\documentclass[preprint,floats,nofootinbib,superscriptaddress,color,showpacs,showkeys]{revtex4}

\usepackage{bm}
\usepackage{amsmath}
\usepackage{graphicx}
\usepackage{graphics}
\usepackage[dvips]{color}
\newcommand{\be}{\begin{equation}} \newcommand{\ee}{\end{equation}}
\newcommand{\ba}{\begin{eqnarray}} \newcommand{\ea}{\end{eqnarray}}

\begin{document}

\title{Nuclear effects in electron reactions and their impact on neutrino
processes}

\pacs{25.30.Pt,  23.40.Bw, 24.10.Jv   }
\keywords{Neutrino scattering, Electron scattering, Superscaling}

\author{Maria B. Barbaro}
\affiliation{Turin University and INFN, Italy}
\author{J.E. Amaro}
\affiliation{University of Granada, Spain}
\author{J.A. Caballero}
\affiliation{University of Sevilla, Spain}
\author{R. Cenni}
\affiliation{INFN, Sezione di Genova, Italy}
\author{T.W. Donnelly}
\affiliation{M.I.T., Boston, USA}
\author{A. Molinari}
\affiliation{Turin University and INFN, Italy}
\author{J.M. Ud\'ias}
\affiliation{Universidad Complutense de Madrid, Spain}


\begin{abstract}
We suggest that superscaling  in electroweak 
interactions with nuclei, namely the observation that the reduced
electron-nucleus cross sections are to a large degree independent of the
momentum transfer and of the nuclear species,
can be used as a tool to obtain precise predictions for
neutrino-nucleus cross sections in both charged and neutral current-induced 
processes.
\end{abstract}


\maketitle

\section{Introduction}

The original idea of scaling in inclusive electron-nucleus scattering 
dates back to 1975, with the seminal work of West \cite{West74}, and was further
developed in Refs. \cite{Alb88,Day90,DS199,DS299}. It basically states that
at high momentum transfer ($q$ larger than about 0.5 GeV/c) 
the ratio between the nuclear and single-nucleon cross sections 
does not depend on two variables ($q$, $\omega$) but only on one
scaling variable ($y$ in a non-relativistic context and $\psi$ in a relativistic
framework). Another type of scaling (the so-called ``second kind scaling'')
concerns the dependence of the above ratio upon the specific target and it is
realized if the latter is inversely proportional to the Fermi
momentum $k_F$. When both kinds of scaling are fulfilled, namely when the 
function 
\begin{equation}
f = k_F \frac{\frac{d\sigma}{d\Omega d\omega}}{Z\sigma_{ep}+
N\sigma_{en}}
\end{equation}
does not vary with the momentum transfer $q$ and with the nuclear target, 
one has ``superscaling'' and $f$ is called superscaling function.
Careful analyses of the $(e,e')$ world data have shown that superscaling
is working to a high degree of accuracy at the left of the quasielastic peak 
(QEP) and that its violations, occurring at the right of the QEP, reside in 
the transverse channel and are mainly (but not only) due to the excitation
of a $\Delta$-resonance. A similar analysis was recently performed in the 
region above the QEP \cite{amaSSM04,Mai09} and it was shown that also here 
superscaling is fulfilled, although to a lesser degree of accuracy, providing 
an appropriate dividing factor and scaling variable, both accounting for 
the $\Delta$-nucleon mass difference, are used. 

As a consequence a good representation of the electromagnetic response can
be obtained 
in both the quasielastic and $\Delta$ regions by embodying the nuclear effects
in two phenomenological superscaling functions, as will be illustrated in the 
next Section. 

The scaling approach can then be inverted and predictions can be made for 
neutrino reactions  by multiplying the superscaling function by the appropriate
single nucleon neutrino factors.
This approach, referred to as ``SuSA'' (SuperScaling Approximation) and
illustrated in Section 3, has the
merit of minimizing the model dependence implicitly associated to any direct
calculation of neutrino-nucleus cross sections, because the complexity of the
nuclear dynamics is extracted by the experimental electron scattering data.
It also has limitations related to the quality of superscaling:
it cannot be applied to very low momentum transfer, where nuclear collective 
effects become important, and it neglects the contribution
of meson exchange currents which, being carried by two-body operators, are
breaking both kinds of scaling. We will shortly comment on these scaling 
violations in Section 4.

\section{The superscaling functions in inclusive electron scattering}

\begin{figure}
\label{fig:scf}
\includegraphics[scale=0.7]{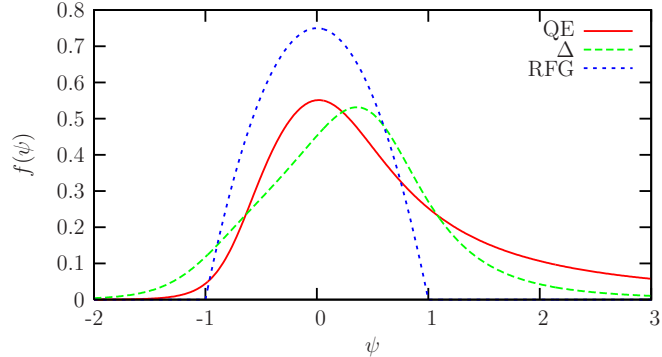}
\caption{The phenomenological QE (red) and $\Delta$ (green) superscaling
functions compared with the RFG result.}
\end{figure}

By applying the procedure outlined in the previous section (see 
Refs.~\cite{DS199,DS299,Jourdan:1996,amaSSM04} for details), the two scaling 
functions shown in Fig.~\ref{fig:scf}, to be used in the QE and $\Delta$
peak regions respectively, can be extracted from the data.
In Fig.~\ref{fig:scf} the Relativistic Fermi Gas (RFG) result, corresponding to the parabola 
$f=\frac{3}{4}(1-\psi^2)$, is also shown for comparison.

\begin{figure}[h]
\label{fig:RMF}
\includegraphics[scale=0.4,angle=270]{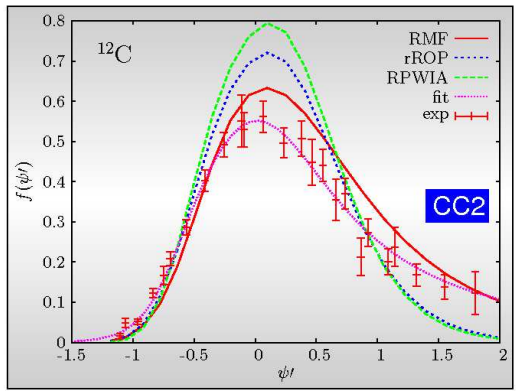}
\caption{
The QE superscaling function $f$ evaluated in different relativistic models 
(see text) and compared with the experimental data. The scaling variable
$\psi^\prime$ includes a small phenomenological energy shift needed in order to
adjust the position of the QEP to the data.
}
\end{figure}

\begin{figure}[h]
\label{fig:BCS}
\includegraphics[scale=0.5]{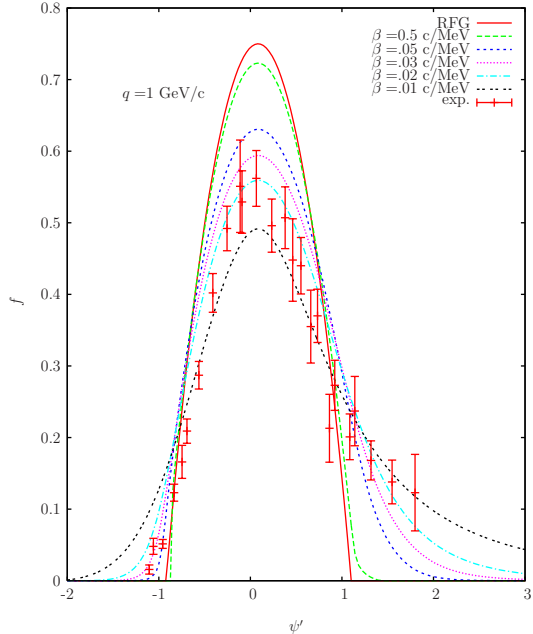}
\caption{
The ``BCS-like'' QE scaling function (see Ref.~\cite{BCS}) evaluated for 
different choices of the parameter $\beta$ controlling the momentum distribution
and compared with the data and the RFG result.
}
\end{figure}

Note that both phenomological superscaling functions 
present an asymmetric shape with a pronounced tail extending into
the region of high transferred energies, corresponding to positive
values of the scaling variable $\psi$. This tail is absent in the RFG result,
which is symmetric in $\psi$. While this asymmetry is
largely absent in most non-relativistic models based on the
impulse approximation, it has been shown that the correct amount of asymmetry
for the QE superscaling function is provided within a few models:
\begin{enumerate}
\item the relativistic impulse approximation when
fnal state interactions (FSI) are described with a relativistic mean field 
(RMF) potential \cite{jac05},
\item a semi-relativistic model including FSI
through a Dirac-equation-based model~\cite{amaFSI},
\item a covariant extension of the relativistic Fermi gas model
which incorporates  correlation effects in nuclei in analogy to the BCS
descriptions of systems of fermions \cite{BCS}.
\end{enumerate}
This is illustrated in Figs.~\ref{fig:RMF} and \ref{fig:BCS}. In Fig.~\ref{fig:RMF} the 
RMF quasielastic scaling function is compared with the data and with other 
relativistic models (the Relativistic Plane Wave Impulse Approximaton and a 
Relativistic Optical 
Model using a real relativistic potential), which are clearly unable to 
reproduce the high energy tail. In Fig.~\ref{fig:BCS} the ``BCS-like'' scaling
function is plotted for different values of a parameter $\beta$, which controls
the slope of the momentum distribution (see Ref.~\cite{BCS} for details): it 
clearly appears that the model has the capability of reproducing the
experimental data providing the nucleon momentum distribution displays a 
sufficiently large high-momentum tail (corresponding to low values of $\beta$).
The connection between the nucleon's momentum distribution and the 
scaling function and its application to neutrino reactions have also been 
explored in Refs.~\cite{anton1,anton2,anton3} within the Coherent Density Fluctuation Model.

The microscopic origin of the difference between the two functions displayed
in Fig.\ref{fig:scf} has been recently explored in Ref.\cite{Mai09}
but relativistic theoretical models for the $\Delta$-resonance region are still
missing, so that the present phenomenological approach is probably the most
reliable one in this kinematical domain.

\begin{figure}[h]
\label{fig:exp}
\includegraphics[scale=0.6,angle=0]{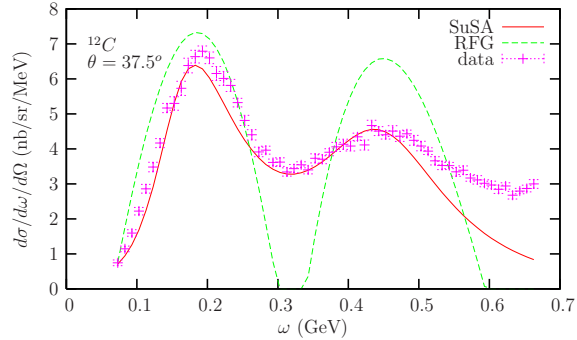}
\caption{
Electron scattering data (from Ref.~\cite{Benhar:2006wy}) compared
with the SuSA (red) and RFG (green) calculations.
}
\end{figure}

With the above ingredients, it is then possible to recalculate for every
nucleus, incident electron energy and scattering angle the inclusive cross
section for  energy transfers $\omega$ below the maximum of the $\Delta$ contribution.
In order to illustrate this, 
in Fig.~\ref{fig:exp} we show
the experimental cross section together with the calculated response 
obtained using the parameterized superscaling functions for a particular
kinematics.
A variety of different kinematics has been explored in Ref. \cite{amaSSM04}
for $^{12}$C and $^{16}$O, since these are the relevant nuclei for the 
MiniBooNE and K2K/T2K neutrino oscillation measurements.
For the data sets which do cover the
$\Delta$ region, typical deviations are 10\% or less, whereas the RFG model clearly
fails to reproduce the data, especially in the ``dip'' region between the two
peaks.

\section{The SuSA model: predictions for neutrino cross sections}

In this Section we present some results obtained within the superscaling
approach for neutrino-nucleus charged and neutral current 
induced processes. 

\subsection{
Charged Current reactions}

According with what previously illustrated we can evaluate the cross sections
relative to the charged current (CC) process, where a lepton $l$ is detected in the 
final state, by multiplying the phenomenological superscaling function
by the corresponding $\nu_l+n\to l^-+p$ cross section. We refer the reader
to Ref.~\cite{amaSSM04} for further details; here we just remind that we use
the Hoehler parameterization~\cite{Hoehler:1976} for the vector form factors
and a dipole axial form factor with a cutoff mass $M_A=1.032$ MeV.

\begin{figure}[h]
\includegraphics[scale=0.5,clip]{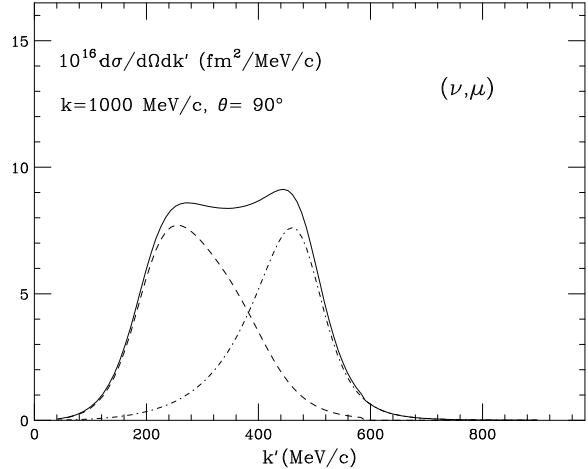}
\label{fig:cc1}
\caption{
Double differential cross section for charged-current neutrino scattering
plotted versus the muon energy for fixed neutrino energy and scattering angle.
The separate $\Delta$ (left peak) and QE (right peak) contributions are 
displayed.
}
\end{figure}

In Fig.~\ref{fig:cc1} we display for illustration the double differential cross
section of neutrino scattering off $^{12}C$ with respect to the outgoing muon 
momentum $k^\prime$ and solid angle $\Omega$ for fixed muon scattering angle 
$\theta=90^o$ and neutrino momentum $k$=1 GeV/c.
The latter is chosen as representative of the kinematics where the scaling 
approach is expected to work well. Results corresponding to different nuclei 
and kinematics as well as to antineutrino scattering can be found in 
Ref.~\cite{amaSSM04}. Note that the quasi-elastic and $\Delta$ peaks have 
roughly the same heighth for this kinematics, although this changes with the
scattering amgle. Note also that the predictions at low momenta $k^\prime$,
to the left of the $\Delta$ peak, are not reliable because our scaling approach
does not fully account for meson production, resonances other than the $\Delta$
and deep inelastic scattering processes. Work aimed to include these 
contributions has been recently performed for electron 
scattering~\cite{Mai09} and will be applied to neutrino scattering in future work.

\begin{figure}[h]
\label{fig:CS_total}
\includegraphics[scale=0.5,clip]{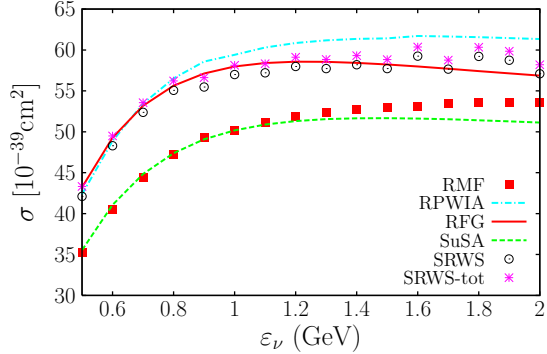}
\caption{
Charged-current neutrino cross section integrated over the muon energy and
the scattering angle versus the neutrino energy (see text for the description of the various curves).
}
\end{figure}

In order to assess the uncertainties related to different nuclear models,
in Fig.~\ref{fig:CS_total}
we plot the fully integrated cross section $\sigma$ as a function of the
incident neutrino energy as calculated within five different 
models~\cite{ama07}: 
the RMF, the Relativistic Plane Wave Impulse Approximation (RPWIA), the RFG,
the SuSA model, and a semi-relativistic shell model~\cite{ama05} including (SRWS-tot)
or not (SRWS) the contribution of the discrete spectrum of the residual
nucleus $^{12}N$ obtained with the Woods-Saxon potential. 
We note that all of the descriptions lead to a similar behavior for the
quasielastic cross section, which increases with the neutrino
energy up to $\varepsilon_\nu\sim$ 1--1.2 GeV and then saturates
to an almost constant value.
However, while the RPWIA, RFG, SRWS and SRWS-tot yield very similar
results, the SuSA and RMF predictions are close to each other and significantly
lower than the RFG. The reduction remains sizeable
even at large $\varepsilon_\nu$ and tends to stabilize,
being of the order of $\sim$15\% at neutrino energies
$\varepsilon_\nu\geq 1.2$ GeV.

Before concluding this Section, a comment is in order concerning the effect of
Pauli blocking (PB). It is well-known that PB,
which is obviously accounted for in the RFG model, only affects
the low momentum and energy transfer region. However in the
process we are considering here this region is kinematically
forbidden due to the mass difference between the initial
($^{12}$C) and residual ($^{12}$N) 
nuclei. As a consequence the effect of PB in the integrated cross
sections turns out to be negligible.

\subsection{
Neutral Current reactions}

For neutral current (NC) reactions, e.g., $^{12}C(\nu,p)\nu X$,
a new feature arises with respect to the CC case. Indeed
in this case the scattered lepton is a neutrino 
and therefore not detected, but the knocked-out nucleon is
detected: it is than the $u$-channel, rather than the $t$-channel, 
whose kinematics are controlled 
(see \cite{Barbaro:1996vd} for discussions of this case).
Accordingly, in the scaling analysis it is not obvious that the
process can be simply related to inclusive electron scattering, 
and therefore to apply the scaling ideas to NC neutrino and antineutrino 
scattering.

\begin{figure}[h]
\label{fig:NC}
\includegraphics[scale=0.6,clip]{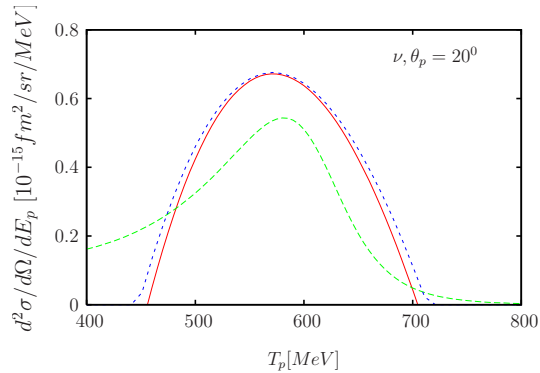}
\caption{Quasielastic differential cross section for neutral
current 1 GeV neutrino scattering from $^{12}$C for proton knockout 
obtained using
the RFG (blue), the factorized approach with the RFG
scaling function (red) and the phenomenological superscaling function
(green).
}
\end{figure}

However it has been proven in Ref.~\cite{amaNC} through a numerical analysis
that the SuSA approach is still applicable. Indeed, as shown in 
Fig.~\ref{fig:NC}, the exact RFG result (blue curve) 
almost coincides with the one obtained under the factorization assumption 
which underlies superscaling (red).
This outcome, closely related to the smooth variation of the single-nucleon
factor inside the integration region, allows us to calculate NC cross sections using 
the SuSA model (green). 
The SuSA cross sections are seen to be lower by about $25\%$ at the peak 
than the RFG results, an effect similar to what was found
for charge-changing processes.
Moreover, the empirical scaling function leads to cross sections
extending both below and above the kinematical region where the
RFG is defined. In particular, the long tail displayed for low
$T_N$-values (corresponding to positive values of the scaling
variable $\psi'$) is noteworthy. This tail arises not only from
the asymmetric shape of the phenomenological scaling function, but
also from the effective NC single-nucleon cross section, which
increases significantly for low $T_N$ values.
%
%
We refer the reader to Ref.~\cite{amaNC} for further results concerning
neutron knockout, antineutrino scattering and the possibility of extracting
information about the possible presence of 
strangeness in the nucleon from neutrino scattering data.

\section{Scaling violations}

Owing to the complexity of nuclear dynamics it is not obvious that
the nuclear response to an electroweak field superscales. 
Indeed several effects are expected to break superscaling to some extent:
off-shellness, collective nuclear excitations, meson-exchange 
currents (MEC), nucleon-nucleon (NN) correlations. To assess the impact of these 
contributions in the QE peak region is then of crucial 
importance.
However a fully relativistic treatment of meson-exchange currents, which is
required by the high energies involved in neutrino experiments, is a very 
demanding task because a large number of diagrams has to be taken into account,
each of them involving multidimensional integrals. Moreover, in order to 
preserve gauge invariance not only the genuine meson exchange currents, where
the virtual boson attaches to the meson, have to be calculated, but also the
associated two-body correlation currents.

Progress in this directions has been made in the last years in a RFG 
framework, where the contribution of the pionic MEC to the QE responses 
have and the associated correlations can be exactly evaluated. 
The full calculation in the 1p-1h sector has been performed in 
Refs.~\cite{ama01,Amaro:2002mj,Amaro:2003yd}, where it was shown that
the MEC break scaling and that they are not negligible, yielding a depletion of
the QEP of about 10-15\%, depending on the kinematics. Moreover, delicate
cancellations occur between the various contributions.
Recently the same calculation has been performed in a semirelativistic shell model
including final state interactions~\cite{ama09}, where it is shown that new effects, 
absent in the RFG case, arise from the interplay between MEC and FSI.

In the 2p-2h sector the evaluation of the MEC contribution is very
demanding from the computational point of view and a complete
gauge invariant calculation is not yet available, although work is in rogress in this direction.
From the results of Refs.~\cite{ADP}, which include the pure MEC but not
the associated correlation diagrams, it appears that these contributions 
are essential if one is to have a quantitative picture of inclusive electron scattering at the
kinematics relevant for neutrino experiments.


\end{document}